# Limitations of Self-Assembly at Temperature 1*

David Doty†, Matthew J. Patitz, and Scott M. Summers‡

Department of Computer Science
Iowa State University
Ames, IA 50011, USA
{ddoty,mpatitz,summers}@cs.iastate.edu

**Abstract**

We prove that if a set $X \subseteq \mathbb{Z}^2$ weakly self-assembles at temperature 1 in a deterministic (Winfree) tile assembly system satisfying a natural condition known as *pumpability*, then $X$ is a finite union of semi-doubly periodic sets. This shows that only the most simple of infinite shapes and patterns can be constructed using pumpable temperature 1 tile assembly systems, and gives evidence for the thesis that temperature 2 or higher is required to carry out general-purpose computation in a tile assembly system. Finally, we show that general-purpose computation *is* possible at temperature 1 if negative glue strengths are allowed in the tile assembly model.

## 1 Introduction

Self-assembly is a bottom-up process by which a small number of fundamental components automatically coalesce to form a target structure. In 1998, Winfree [10] introduced the (abstract) Tile Assembly Model (TAM) – an "effectivization" of Wang tiling [8, 9] – as an over-simplified mathematical model of the DNA self-assembly pioneered by Seeman [7]. In the TAM, the fundamental components are un-rotatable, but translatable square "tile types" whose sides are labeled with glue "colors" and "strengths." Two tiles that are placed next to each other *interact* if the glue colors on their abutting sides match, and they *bind* if the strength on their abutting sides matches with total strength at least a certain ambient "temperature," usually taken to be 1 or 2.

Despite its deliberate over-simplification, the TAM is a computationally and geometrically expressive model at temperature 2. The reason is that, at temperature 2, certain tiles are not permitted to bond until *two* tiles are already present to match the glues on the bonding sides, which enables cooperation between different tile types to control the placement of new tiles. Winfree [10] proved that at temperature 2 the TAM is computationally universal and thus can be directed algorithmically.

In contrast, we aim to prove that the lack of cooperation at temperature 1 inhibits the sort of complex behavior possible at temperature 2. Our main theorem concerns the *weak self-assembly*

*This research was supported in part by National Science Foundation grants 0652569 and 0728806.

†This author's research was partially supported by NSF grant CCF:0430807.

‡This author's research was supported in part by NSF-IGERT Training Project in Computational Molecular Biology Grant number DGE-0504304.

of subsets of $\mathbb{Z}^2$ using temperature 1 tile assembly systems. Informally, a set $X \subseteq \mathbb{Z}^2$ weakly self-assembles in a tile assembly system $\mathcal{T}$ if some of the tile types of $\mathcal{T}$ are painted black, and $\mathcal{T}$ is guaranteed to self-assemble into an assembly $\alpha$ of tiles such that $X$ is precisely the set of integer lattice points at which $\alpha$ contains black tile types. As an example, Winfree [10] exhibited a temperature 2 tile assembly system, directed by a clever XOR-like algorithm, that weakly self-assembles a well-known set, the discrete Sierpinski triangle, onto the first quadrant. Note that the underlying *shape* (set of all points containing a tile, whether black or not) of Winfree's construction is an infinite canvas that covers the entire first quadrant, onto which a more sophisticated set, the discrete Sierpinski triangle, is painted.

We show that under a plausible assumption, temperature 1 tile systems weakly self-assemble only a limited class of sets. To prove our main result, we require the hypothesis that the tile system is *pumpable*. Informally, this means that every sufficiently long path of tiles in an assembly of this system contains a segment in which the same tile type repeats (a condition clearly implied by the pigeonhole principle), and that furthermore, the subpath between these two occurrences can be repeated indefinitely ("pumped") along the same direction as the first occurrence of the segment, without "colliding" with a previous portion of the path. We give a counterexample in Section 3 (Figure 1) of a path in which the same tile type appears twice, yet the segment between the appearances cannot be pumped without eventually resulting in a collision that prevents additional pumping. The hypothesis of pumpability states (roughly) that in every sufficiently long path, despite the presence of some repeating tiles that cannot be pumped, *there exists* a segment in which the same tile type repeats that *can* be pumped. In the above-mentioned counterexample, the paths constructed to create a blocked segment always contain some previous segment that *is* pumpable. We conjecture that this phenomenon, pumpability, occurs in every temperature 1 tile assembly system that produces a unique infinite structure. We discuss this conjecture in greater detail in Section 6.

A *semi-doubly periodic* set $X \subseteq \mathbb{Z}^2$ is a set of integer lattice points with the property that there are three vectors $\vec{b}$ (the "base point" of the set), $\vec{u}$, and $\vec{v}$ (the two periods of the set), such that $X = \left\{ \vec{b} + n \cdot \vec{u} + m \cdot \vec{v} \;\middle|\; n, m \in \mathbb{N} \right\}$. That is, a semi-doubly periodic set is a set that repeats infinitely along two vectors (linearly independent vectors in the non-degenerate case), starting at some base point $\vec{b}$. We show that any directed, pumpable, temperature 1 tile assembly system weakly self-assembles a set $X \subseteq \mathbb{Z}^2$ that is a finite union of semi-doubly periodic sets.

It is our contention that weak self-assembly captures the intuitive notion of what it means to "compute" with a tile assembly system. For example, the use of tile assembly systems to build shapes is captured by requiring all tile types to be black, in order to ask what set of integer lattice points contain any tile at all (so-called *strict self-assembly*). However, weak self-assembly is a more general notion. For example, Winfree's above mentioned result shows that the discrete Sierpinski triangle weakly self-assembles at temperature 2 [6], yet this shape does not strictly self-assemble at *any* temperature [2]. Hence weak self-assembly allows for a more relaxed notion of set building, in which intermediate space can be used for computation, without requiring that the space filled to carry out the computation also represent the final result of the computation.

As another example, there is a standard construction [10] by which a single-tape Turing machine may be simulated by a temperature 2 tile assembly system. Regardless of the semantics of the Turing machine (whether it decides a language, enumerates a language, computes a function, etc.), it is routine to represent the result of the computation by the weak self-assembly of some set. For example, Patitz and Summers [3] showed that for any decidable language $A \subseteq \mathbb{N}$, $A$'s projection along the $X$-axis $\left(\text{the set}\left\{\ (x,0)\in\mathbb{N}^2\ \middle|\ x\in A\ \right\}\right)$ weakly self-assembles in a temperature 2 tile assembly system. As another example, if a Turing machine computes a function $f:\mathbb{N}\to\mathbb{N}$, it is routine to design a tile assembly system based on Winfree's construction such that, if the seed assembly is used to encode the binary representation of a number $n\in\mathbb{N}$, then the tile assembly system weakly self-assembles the set

$$\left\{\ (k,0)\in\mathbb{N}^2\ \middle|\ \begin{array}{l}\text{the } k^{\text{th}} \text{ least significant bit of the}\\ \text{binary representation of } f(n) \text{ is } 1\end{array}\right\}.$$

Our result is motivated by the thesis that if a tile assembly system can reasonably be said to "compute", then the result of this computation can be represented in a straightforward manner as a set $X\subseteq\mathbb{Z}^2$ that weak self-assembles in the tile assembly system, or a closely related tile assembly system. Our examples above provide evidence for this thesis, although it is as informal and unprovable as the Church-Turing thesis. On the basis of this claim, and the triviality of semi-doubly periodic sets (shown more formally in Observation 4.2), we conclude that our main result implies that directed, pumpable, temperature 1 tile assembly systems are incapable of general-purpose deterministic computation, without further relaxing the model.

This paper is organized as follows. Section 2 provides background definitions and notation for the abstract TAM. Section 3 introduces new definitions and concepts that are required in proving our main theorem. Section 4 states our main theorem, and contains an observation justifying the suggestion that semi-doubly periodic sets are computationally very simple, based on their relationship to regular languages. Section 5 shows an application of our theorem, showing that, unlike the case of temperature 2, no non-trivial discrete self-similar fractal – such as the discrete Sierpinski triangle – weakly self-assembles at temperature 1 in a directed, pumpable tile assembly system. Section 6 concludes the paper and discusses open questions. Section 6 also observes that a relaxation of the tile assembly model, allowing negative glue strengths and allowing glues with different colors to interact, *is* capable of general-purpose computation. Section 7 is a technical appendix containing proofs of some results.

## 2 The Abstract Tile Assembly Model

We work in the 2-dimensional discrete space $\mathbb{Z}^2$. Define the set $U_2=\{(0,1),(1,0),(0,-1),(-1,0)\}$ to be the set of all *unit vectors*, i.e., vectors of length 1 in $\mathbb{Z}^2$. We write $[X]^2$ for the set of all 2-element subsets of a set $X$. All *graphs* here are undirected graphs, i.e., ordered pairs $G=(V,E)$, where $V$ is the set of *vertices* and $E\subseteq[V]^2$ is the set of *edges*.

We now give a brief and intuitive sketch of the Tile Assembly Model that is adequate for reading this paper. More formal details and discussion may be found in [2, 4, 5, 10]. Our notation is that of [2], which contains a self-contained introduction to the Tile Assembly Model for the reader unfamiliar with the model.

Intuitively, a tile type $t$ is a unit square that can be translated, but not rotated, having a well-defined "side $\vec{u}$" for each $\vec{u}\in U_2$. Each side $\vec{u}$ of $t$ has a "glue" of "color" $\text{col}_t(\vec{u})$ – a string over some fixed alphabet $\Sigma$ – and "strength" $\text{str}_t(\vec{u})$ – a nonnegative integer – specified by its type $t$. Two tiles $t$ and $t'$ that are placed at the points $\vec{a}$ and $\vec{a}+\vec{u}$ respectively, *bind* with *strength* $\text{str}_t(\vec{u})$ if and only if $(\text{col}_t(\vec{u}),\text{str}_t(\vec{u}))=(\text{col}_{t'}(-\vec{u}),\text{str}_{t'}(-\vec{u}))$.

Given a set $T$ of tile types, an *assembly* is a partial function $\alpha:\mathbb{Z}^2\dashrightarrow T$, with points $\vec{x}\in\mathbb{Z}^2$ at which $\alpha(\vec{x})$ is undefined interpreted to be empty space, so that dom $\alpha$ is the set of points with tiles. $\alpha$ is *finite* if $|\text{dom } \alpha|$ is finite. For assemblies $\alpha$ and $\alpha'$, we say that $\alpha$ is a *subconfiguration* of $\alpha'$, and write $\alpha \sqsubseteq \alpha'$, if $\text{dom } \alpha \subseteq \text{dom } \alpha'$ and $\alpha(\vec{x}) = \alpha'(\vec{x})$ for all $x \in \text{dom } \alpha$.

Let $\alpha$ be an assembly and $B \subseteq \mathbb{Z}^2$. *$\alpha$ restricted to $B$*, written as $\alpha \upharpoonright B$, is the unique assembly satisfying $(\alpha \upharpoonright B) \sqsubseteq \alpha$, and $\text{dom } (\alpha \upharpoonright B) = B$. If $\pi$ is a sequence over $\mathbb{Z}^2$ (such as a path), then we write $\alpha \upharpoonright \pi$ to mean $\alpha$ restricted to the set of points in $\pi$. If $A \subseteq \text{dom } \alpha$, we write $\alpha \setminus A = \alpha \upharpoonright (\text{dom } \alpha - A)$. If $\vec{0} \neq \vec{v} \in \mathbb{Z}^2$, then the *translation of $\alpha$ by $\vec{v}$* is defined as the assembly $(\alpha + \vec{v})$ satisfying, for all $\vec{a} \in \mathbb{Z}^2$,

$$(\alpha + \vec{v})(\vec{a}) = \begin{cases} \alpha(\vec{a}) & \text{if } \vec{a} - \vec{v} \in \text{dom } \alpha \\ \uparrow & \text{otherwise.} \end{cases}$$

A *grid graph* is a graph $G = (V, E)$ in which $V \subseteq \mathbb{Z}^2$ and every edge $\{\vec{a}, \vec{b}\} \in E$ has the property that $\vec{a} - \vec{b} \in U_2$. The *binding graph of* an assembly $\alpha$ is the grid graph $G_\alpha = (V, E)$, where $V = \text{dom } \alpha$, and $\{\vec{m}, \vec{n}\} \in E$ if and only if (1) $\vec{m} - \vec{n} \in U_2$, (2) $\text{col}_{\alpha(\vec{m})}(\vec{n} - \vec{m}) = \text{col}_{\alpha(\vec{n})}(\vec{m} - \vec{n})$, and (3) $\text{str}_{\alpha(\vec{m})}(\vec{n} - \vec{m}) > 0$. An assembly is *$\tau$-stable*, where $\tau \in \mathbb{N}$, if it cannot be broken up into smaller assemblies without breaking bonds of total strength at least $\tau$ (i.e., if every cut of $G_\alpha$ cuts edges, the sum of whose strengths is at least $\tau$).

Self-assembly begins with a *seed assembly* $\sigma$ (typically assumed to be finite and $\tau$-stable) and proceeds asynchronously and nondeterministically, with tiles adsorbing one at a time to the existing assembly in any manner that preserves stability at all times.

A *tile assembly system* (*TAS*) is an ordered triple $\mathcal{T} = (T, \sigma, \tau)$, where $T$ is a finite set of tile types, $\sigma$ is a seed assembly with finite domain, and $\tau$ is the temperature. In subsequent sections of this paper, we assume that $\tau = 1$ unless explicitly stated otherwise. An *assembly sequence* in a TAS $\mathcal{T} = (T, \sigma, 1)$ is a (possibly infinite) sequence $\vec{\alpha} = (\alpha_i \mid 0 \leq i < k)$ of assemblies in which $\alpha_0 = \sigma$ and each $\alpha_{i+1}$ is obtained from $\alpha_i$ by the "$\tau$-stable" addition of a single tile. The *result* of an assembly sequence $\vec{\alpha}$ is the unique assembly $\text{res}(\vec{\alpha})$ satisfying $\text{dom } \text{res}(\vec{\alpha}) = \bigcup_{0 \leq i < k} \text{dom } \alpha_i$ and, for each $0 \leq i < k$, $\alpha_i \sqsubseteq \text{res}(\vec{\alpha})$.

Note that, if $\mathcal{T} = (T, \sigma, 1)$, then there is a finite path from the seed to a point $\vec{x} \in \text{dom } \alpha$, denoted as $\pi_{\vec{0},\vec{x}}$ if and only if there is an assembly sequence $\vec{\alpha}$ satisfying $\text{res}(\vec{\alpha}) = \alpha \upharpoonright \pi_{\vec{0},\vec{x}}$.

We write $\mathcal{A}[\mathcal{T}]$ for the *set of all producible assemblies of $\mathcal{T}$*. An assembly $\alpha$ is *terminal*, and we write $\alpha \in \mathcal{A}_\Box[\mathcal{T}]$, if no tile can be stably added to it. We write $\mathcal{A}_\Box[\mathcal{T}]$ for the *set of all terminal assemblies of $\mathcal{T}$*. A TAS $\mathcal{T}$ is *directed*, or *produces a unique assembly*, if it has exactly one terminal assembly i.e., $|\mathcal{A}_\Box[\mathcal{T}]| = 1$. The reader is cautioned that the term "directed" has also been used for a different, more specialized notion in self-assembly [1]. We interpret "directed" to mean "deterministic", though there are multiple senses in which a TAS may be deterministic or nondeterministic.

A set $X \subseteq \mathbb{Z}^2$ *weakly self-assembles* if there exists a TAS $\mathcal{T} = (T, \sigma, 1)$ and a set $B \subseteq T$ ($B$ constitutes the "black" tiles) such that $\alpha^{-1}(B) = X$ holds for every assembly $\alpha \in \mathcal{A}_\Box[\mathcal{T}]$. A set $X$ *strictly self-assembles* if there is a TAS $\mathcal{T}$ for which every assembly $\alpha \in \mathcal{A}_\Box[\mathcal{T}]$ satisfies $\text{dom } \alpha = X$. Note that if $X$ strictly self-assembles, then $X$ weakly self-assembles. (Let all tiles be black.)

## 3 Pumpability, Finite Closures, and Combs

Throughout this section, let $\mathcal{T} = (T, \sigma, 1)$ be a directed TAS, and $\alpha$ be the unique assembly satisfying $\alpha \in \mathcal{A}_\Box[\mathcal{T}]$. Further, we assume without loss of generality that $\sigma$ consists of a single "seed" tile type placed at the origin.

Given, $\vec{0} \neq \vec{v} \in \mathbb{Z}^2$, a *$\vec{v}$-semi-periodic path in $\alpha$ originating at $\vec{a}_0 \in$ dom $\alpha$* is an infinite, simple path $\pi = (\vec{a}_0, \vec{a}_1, \ldots)$ in the binding graph $G_\alpha$ such that there is a constant $k \in \mathbb{N}$ such that, for all $j \in \mathbb{N}$, $\pi[j+k] = \pi[j] + \vec{v}$, and $\alpha(\pi[j+k]) = \alpha(\pi[j])$. Intuitively, $\vec{v}$ is the "geometric" period of the path – the straight-line vector between two repeating tile types – while $k$ is the "linear" period – the number of tiles that must be traversed along the path before the tile types repeat, which is at least $||\vec{v}||_1$, but possibly larger if the segment from $\pi[j]$ to $\pi[j] + \vec{v}$ is "winding".

An *eventually $\vec{v}$-semi-periodic path in $\alpha$ originating at $\vec{a}_0 \in$ dom $\alpha$* is an infinite, simple path $\pi = (\vec{a}_0, \vec{a}_1, \ldots)$ in the binding graph $G_\alpha$ for which there exists $s \in \mathbb{N}$ such that the path $\pi' = (\pi[s], \pi[s+1], \ldots)$ is a $v$-semi-periodic path in $\alpha$ originating at $\pi[s]$. Let the *initial segment length* be the smallest index $s^*$ such that $\pi' = (\pi[s^* - k], \pi[s^* - k + 1], \ldots)$ is a $\vec{v}$-semi-periodic path originating at the point $\pi[s^* - k]$. The *initial segment of* $\pi$ is the path $\pi[0 \ldots i^* - 1]$ (for technical reasons, we enforce the initial segment of $\pi$ to contain the simple path $\pi[0 \ldots s]$ along with one period of $\pi'$). The *tail of* $\pi$ is $\pi[s^* \ldots]$. Note that the tail of an eventually $\vec{v}$-semi-periodic path is simply a $\vec{v}$-semi-periodic path originating at $\pi[s^*]$. An *eventually $\vec{v}$-periodic path in $\alpha$* is an eventually $\vec{v}$-periodic path in $\alpha$ originating at $\vec{a}_0$ for some $\vec{a}_0 \in$ dom $\alpha$. We say that $\pi$ is a *$\vec{v}$-periodic* path in $\alpha$ if $\pi = (\ldots, \vec{a}_{-1}, \vec{a}_0, \vec{a}_1, \ldots)$ is a two-way infinite simple path such that, for all $j \in \mathbb{Z}$, $\alpha(\pi[j] + \vec{v}) = \alpha(\pi[j])$.

Let $\vec{w}, \vec{x} \in$ dom $\alpha$, $\pi$ be a simple path from $\vec{w}$ to $\vec{x}$ in the binding graph $G_\alpha$, and $i, k \in \mathbb{N}$ with $0 \leq i < k \leq |\pi|$. We say that $\pi$ has a *pumpable segment* $\pi[i \ldots k]$ (with respect to $\vec{v} = \pi[k] - \pi[i]$) if there exists a $\vec{v}$-semi-periodic path $\pi'$ in $\alpha$ originating at $\pi[i]$ and $\pi'[0 \ldots k - i] = \pi[i \ldots k]$.

Intuitively, the path $\pi$ has a pumpable segment if, after some initial sequence of tile types, it consists of a subsequence of tile types which is repeated in the same direction an infinite number of times, one after another. Figure 1 shows an assembly in which the same tile type repeats along

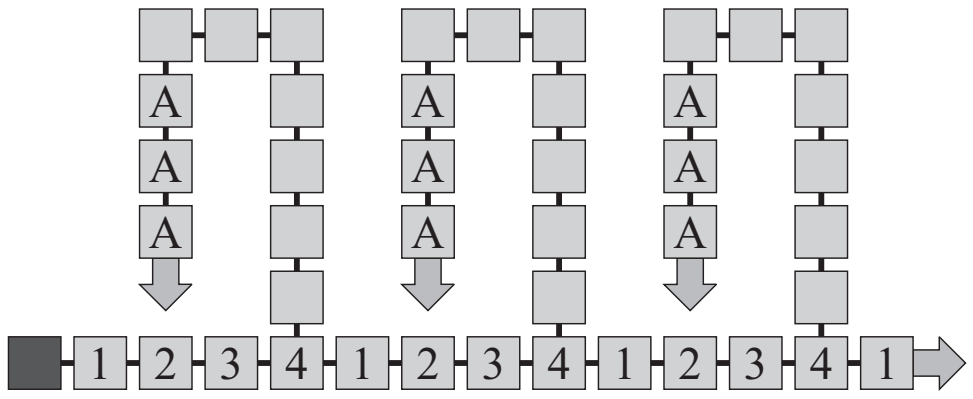

Figure 1: An assembly containing a path with repeating tiles A-A that do *not* form a pumpable segment, because they are blocked from infinite growth by the existing assembly. Note, however, that any sufficiently long path from the origin (at the left) contains a pumpable segment, namely the repeating segment 1-2-3-4-1 along the bottom row, which can be pumped infinitely to the right.

a path, but the segment between the occurrences is not pumpable.

Let $c \in \mathbb{N}$ and $\vec{v} \in \mathbb{Z}^2$. The *diamond of radius $c$ centered about the point $\vec{v}$* is the set of points defined as $D(c, \vec{v}) = \{\ (x, y) + \vec{v} \mid |x| + |y| \leq c\ \}$. Let $\vec{w}, \vec{x} \in$ dom $\alpha$, $c \in \mathbb{N}$, and $\pi$ be a simple path from $\vec{w}$ to $\vec{x}$ in the binding graph $G_\alpha$. We say that $\pi$ is a *pumpable path from $\vec{w}$ to $\vec{x}$ in $\alpha$* if it contains a pumpable segment $\pi[i \ldots k]$ for some $i, k \in \mathbb{N}$ such that $0 \leq i < k \leq |\pi|$. We say that $\mathcal{T}$ is *$c$-pumpable* if there exists $c \in \mathbb{N}$ such that for every $\vec{w}, \vec{x} \in$ dom $\alpha$ with $\vec{x} \notin D(c, \vec{w})$, there exists a pumpable path $\pi$ from $\vec{x}$ to $\vec{w}$ in $\alpha$.

Figure 2 shows, from left to right, (1) a partially complete assembly beginning from the (dark grey) seed tile, where the dark notches between adjacent tiles represent strength 1 bonds, and a tile selected for the example, (2) the full path leading from the seed to the selected tile, (3) the tile types for a segment of the path, showing the repeating pattern of tile types '1-2-3-4', and (4) an extended version of the path which shows its ability to be pumped.

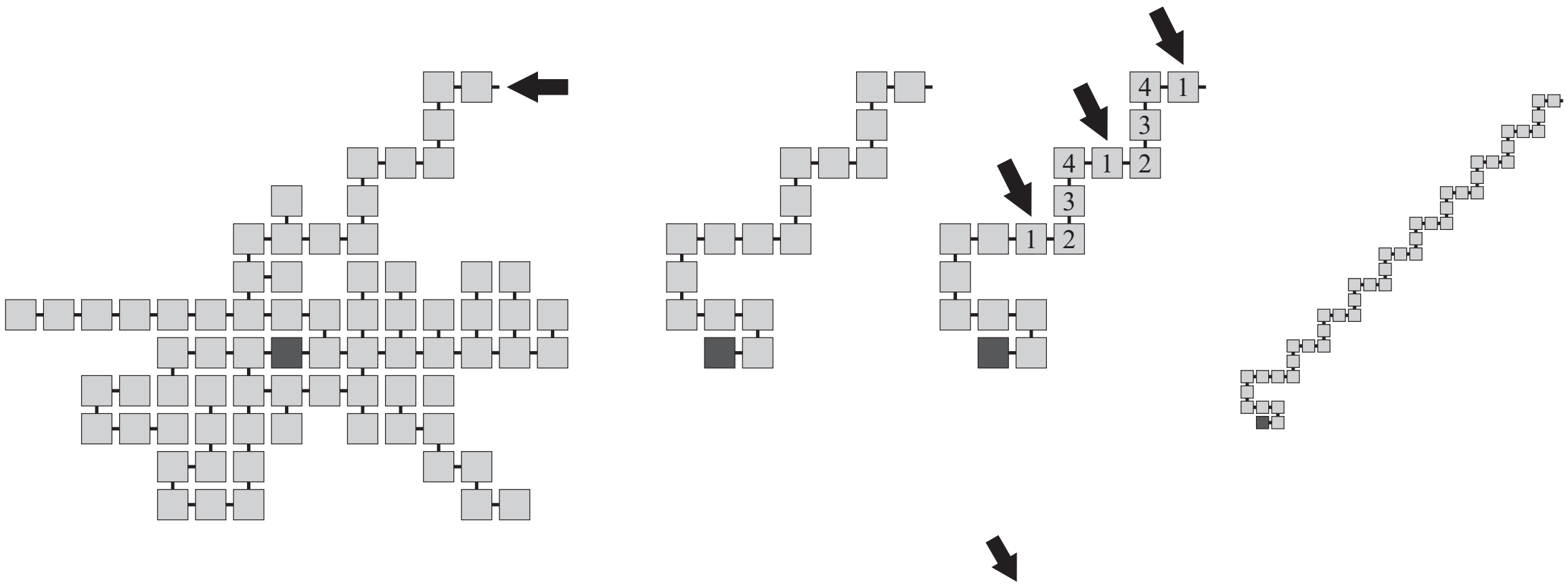

Figure 2: A partial assembly, selected path, pumpable segment, and pumpable path

In our proof, it is helpful to consider extending an assembly in such a way that no individual tile in the existing assembly is extended by more than a finite amount (though an infinite assembly may have an infinite number of tiles that can each be extended by a finite amount). We call such an extension the *finite closure* of the assembly, and define it formally as follows. Let $\alpha' \in \mathcal{A}[\mathcal{T}]$. We say that the *finite closure* of $\alpha'$ is the unique assembly $\mathcal{F}(\alpha')$ satisfying

1. $\alpha' \sqsubseteq \mathcal{F}(\alpha')$, and

2. dom $\mathcal{F}(\alpha')$ is the set of all points $\vec{x} \in \mathbb{Z}^2$ such that every infinite simple path in the binding graph $G_\alpha$ containing $\vec{x}$ intersects dom $\alpha'$.

Intuitively, this means that if we extend $\alpha'$ by only those "portions" that will eventually stop growing, the finite closure is the super-assembly that will be produced. That is, any attempt to "leave" $\alpha'$ through the finite closure and go infinitely far will eventually run into $\alpha'$ again. If $\alpha'$ is terminal, then $\alpha'$ is its own finite closure. Note that in general, the finite closure of an assembly $\alpha'$ is not the result of adding finitely many tiles to $\alpha'$. For instance, if infinitely many points of $\alpha'$ allow exactly one tile to be added, the finite closure adds infinitely many points to $\alpha'$. However, the finite closure of a finite assembly is always a finite assembly.

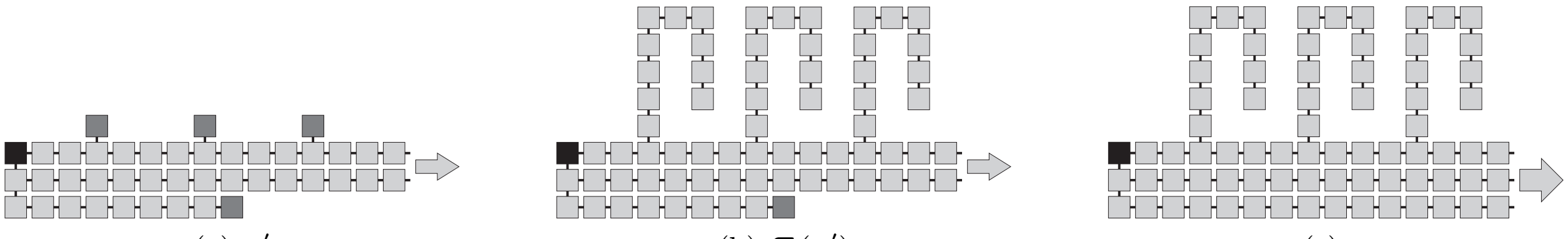

(a) $\alpha'$ (b) $\mathcal{F}(\alpha')$ (c) $\alpha$

Figure 3: Example of a finite closure. The dark gray points represent locations at which tiles can attach.

For an example of a finite closure of an assembly, see Figure 3. Figure 3c shows the terminal assembly which consists of three rows of tiles that continue infinitely to the right (denoted by the arrow), with a 10 tile upward projection occurring at every fourth column. Figure 3a shows an assembly $\alpha'$ which consists of two rows of tiles continuing infinitely to the right but with an incomplete bottom row and none of the full upward projections. Figure 3b shows $\mathcal{F}(\alpha')$, which is the finite closure of $\alpha'$. Notice that an infinite number of tiles were added to $\alpha'$ to create $\mathcal{F}(\alpha')$ since an infinite number of upward projections were added, each consisting of only 10 tiles. However, also note that the bottom row is not grown because that row consists of an infinite path and therefore cannot be part of the finite closure.

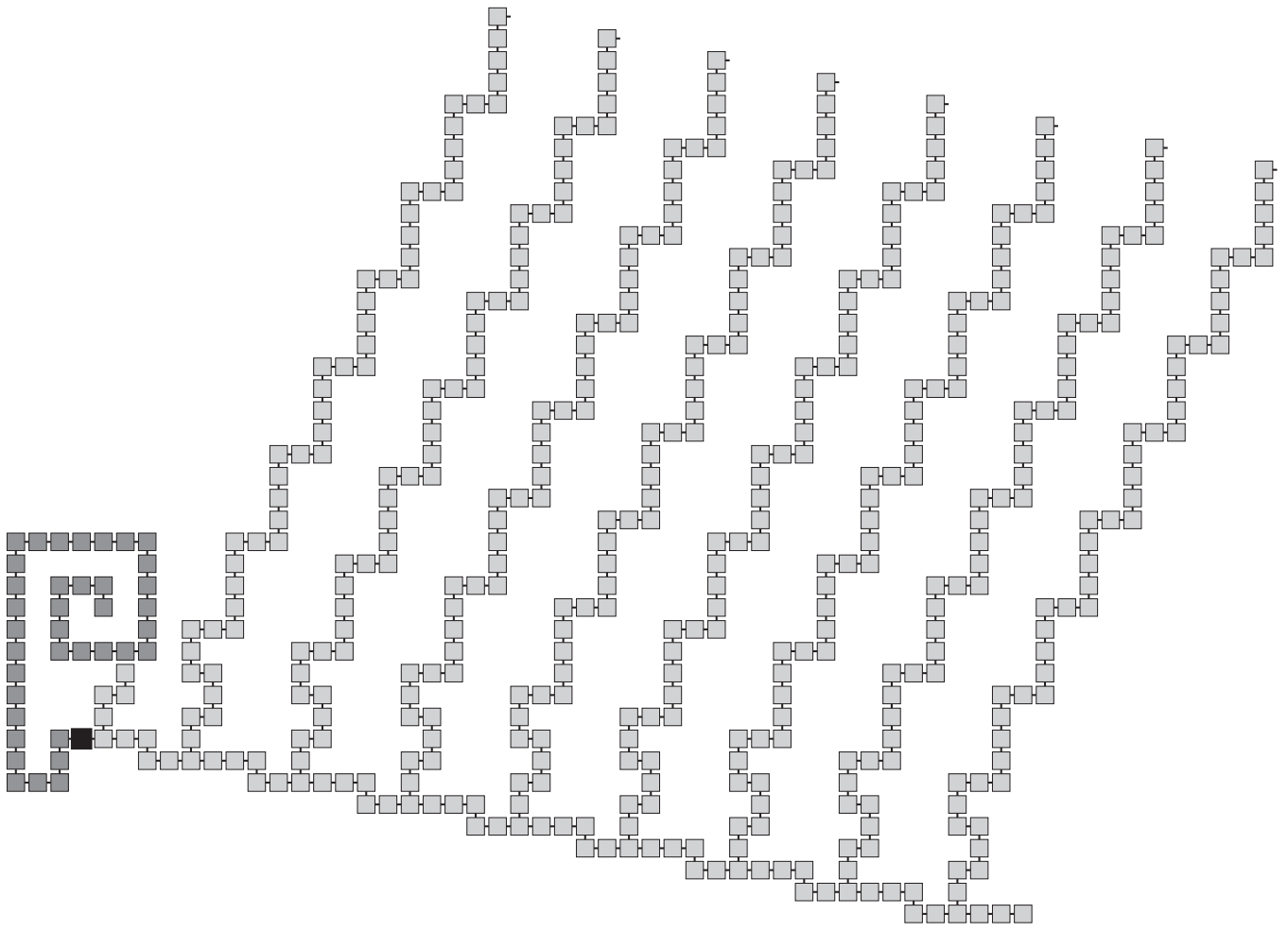

Figure 4: Example of a comb connected to a hard-coded assembly (dark tiles that spiral). The seed is the tile at the center of the spiral. The comb originates at the bottom most black tile. Note that a finite number of "attempts at teeth" (one in the above example) may be blocked before the infinite teeth are allowed to grow.

Let $\pi^{\rightarrow}$ be an eventually $\vec{v}$-semi-periodic path in $\alpha$ originating at $\vec{0}$ where $\vec{0} \neq \vec{v} \in \mathbb{Z}^2$. Let $\vec{u} \in \mathbb{Z}^2$ such that $\vec{u} \neq z \cdot \vec{v}$ for all $z \in \mathbb{R}$. Suppose that there is a point $\vec{b}$ on the tail of $\pi^{\rightarrow}$ such that there is an eventually $\vec{u}$-semi-periodic path $\pi^{\uparrow}$ in $\alpha$ originating at $\vec{b}$ such that $\pi^{\rightarrow} \cap \pi^{\uparrow} = \left\{\vec{b}\right\}$. Define the assembly $\alpha^* = \alpha \upharpoonright \left(\pi^{\rightarrow} \cup \bigcup_{n \in \mathbb{N}} \left(\pi^{\uparrow} + n \cdot \vec{v}\right)\right)$. It is easy to see that $\alpha^*$ need not be a producible assembly. We say that $\alpha^*$ is a *comb in* (with respect to $\pi^{\rightarrow}$ and $\pi^{\uparrow}$) if, for every $n \in \mathbb{N}$, $\alpha^* \upharpoonright \pi^{\uparrow} + n \cdot \vec{v} = \alpha^* \upharpoonright \left(\pi^{\uparrow} + n \cdot \vec{v}\right)$. We refer to the assembly $\alpha^* \upharpoonright \pi$ as the *base* of $\alpha^*$. For any $n \in \mathbb{N}$, we define the $n^{th}$ *tooth of* $\alpha^*$ to be the assembly $\alpha \upharpoonright \left(\pi^{\uparrow} + n \cdot \vec{v}\right)$. We say that the comb $\alpha^*$ *starts at the point* $\pi[s]$ (the point at which the eventually-periodic path $\pi^{\rightarrow}$ becomes periodic). It follows from the definition that, if $\alpha^*$ is a comb in $\alpha$, then $\alpha^* \in \mathcal{A}[\mathcal{T}]$.

See Figure 4 for an example of a comb. A comb, intuitively, is a generalization of the assembly in which an infinite periodic one-way path (the base) grows along the positive $x$-axis, and once per period, an infinite periodic path (a tooth) grows in the positive $y$ direction, creating an infinite number of "teeth". The generalizations are that (1) the base and teeth need not run parallel to either axis, and (2), the teeth may have some initial hard-coded tiles before the repeating periodic segment begins. Also, it is possible at temperature 1 to build multiple combs with the same base, but with different teeth, growing in either direction.

## 4 Main Result

We show in this section that only "simple" sets weakly self-assemble in directed, pumpable tile assembly systems at temperature 1. We now formally define "simple."

**Definition 4.1.** A set $X \subseteq \mathbb{Z}^2$ is *semi-doubly periodic* if there exist three vectors $\vec{b}$, $\vec{u}$, and $\vec{v}$ such that

$$X = \left\{ \left. \vec{b} + n \cdot \vec{u} + m \cdot \vec{v} \;\right|\; n, m \in \mathbb{N} \right\}.$$

Note that a set that is periodic along only one dimension is also semi-doubly periodic, by our definition, since this corresponds to the condition that exactly one of the vectors $\vec{u}$ or $\vec{v}$ is equal to $\vec{0}$ (or if $\vec{u} = \vec{v}$). Similarly, if $\vec{u} = \vec{v} = \vec{0}$, then the definition of semi-doubly periodic is equivalent to $A$ being a singleton set. The following observation justifies the intuition that finite unions of semi-doubly periodic sets constitute only the computationally simplest subsets of $\mathbb{Z}^2$.

**Observation 4.2.** *Let $A \subseteq \mathbb{Z}^2$ be a finite union of semi-doubly periodic sets. Then the unary languages $L_{A,x} = \left\{ \left. 0^{|x|} \;\right|\; (x, y) \in A \text{ for some } y \in \mathbb{Z} \right\}$ and $L_{A,y} = \left\{ \left. 0^{|y|} \;\right|\; (x, y) \in A \text{ for some } x \in \mathbb{Z} \right\}$ consisting of the unary representations of the projections of $A$ onto the $x$-axis and $y$-axis, respectively, are regular languages.*

A proof of Observation 4.2 is included in the technical appendix. The following theorem is the main result of this paper.

**Theorem 4.3.** *Let $\mathcal{T} = (T, \sigma, 1)$ be a directed, pumpable TAS. If a set $X \subseteq \mathbb{Z}^2$ weakly self-assembles in $\mathcal{T}$, then $X$ is a finite union of semi-doubly periodic sets.*

A proof of Theorem 4.3 is given in the technical appendix. The proof idea of Theorem 4.3 is as follows. Suppose that $\alpha \in \mathcal{A}_{\Box}[\mathcal{T}]$. Note that $\alpha$ is unique since $\mathcal{T}$ is directed. Either $\alpha$ is an infinite "grid" that fills the plane, or there exists finitely many semi-doubly periodic combs and semi-periodic paths that, taken together, "cover" every point in $X$ (in the sense that each such point is in the finite closure of one of these combs or paths).

The reason for this is that each comb is defined by two vectors $\vec{u}$ (the base) and $\vec{v}$ (the teeth), and these vectors form a "basis" for the space of points located within the cone formed by the base and the first tooth of the comb. While the vectors do not reach every point in this cone, they reach within a constant distance of every point in the cone, and the doubly periodic regularity of the teeth and base enforces doubly periodic regularity in between the teeth as well. Of course, not all combs have teeth, in which case the comb is just a periodic path.

We associate each black tile with some periodic path or comb that begins in a fixed radius about the origin (utilizing the fact that a path cannot go far from the origin without having pumpable segments that can be used to construct a periodic path or comb). The finite number of combs and periodic paths originating within this radius tells us that the number of semi-doubly periodic sets of (locations tiled by) black tiles that they each define is finite.

## 5 An Application to Discrete Self-Similar Fractals

In this section, we use Theorem 4.3 to show that no discrete self-similar fractal weakly self-assembles in any temperature 1 tile assembly system that is pumpable and directed. Since Winfree [10] showed that one particular discrete self-similar fractal, the discrete Sierpinkski triangle, self-assembles at temperature 2, this provides a concrete example of computation that is possible (and simple) at temperature 2, but impossible at temperature 1, assuming directedness and pumpability.

**Definition 5.1.** Let $1 < c \in \mathbb{N}$, and $X \subsetneq \mathbb{N}^2$ (we do not consider $\mathbb{N}^2$ to be a self-similar fractal). We say that $X$ is a *$c$-discrete self-similar fractal*, if there is a set $V \subseteq \{0, \ldots, c-1\} \times \{0, \ldots, c-1\}$ with

$$V \notin \{\{(i,i) \mid 0 \leq i < c\}, \{(i,0) \mid 0 \leq i < c\}, \{(0,i) \mid 0 \leq i < c\}\}$$

such that

$$X = \bigcup_{i=0}^{\infty} X_i,$$

where $X_i$ is the $i^{\text{th}}$ *stage* satisfying $X_0 = \{(0,0)\}$, and $X_{i+1} = X_i \cup \left(X_i + c^i V\right)$. In this case, we say that $V$ *generates* $X$.

**Definition 5.2.** $X \subsetneq \mathbb{N}^2$. We say that $X$ is a *discrete self-similar fractal* if it is a $c$-discrete self-similar fractal for some $c \in \mathbb{N}$.

The following observation is clear by Definition 5.1.

**Observation 5.3.** *If $X \subsetneq \mathbb{N}^2$ is a discrete self-similar fractal then $X$ is not a finite union of doubly periodic sets.*

**Theorem 5.4.** *Let $X \subsetneq \mathbb{N}^2$ be a discrete self-similar fractal.*

1. *$X$ does not weakly self-assemble in any tile assembly system that is pumpable and directed.*
2. *$X$ does not strictly self-assemble in any tile assembly system that is pumpable and directed.*

*Proof.* (1) follows directly from Observation 5.3. To prove (2), note that if a set $X \subseteq \mathbb{Z}^2$ weakly self-assembles then, by definition, $X$ strictly self-assembles. $\square$

# 6 Conclusion

We have studied the class of shapes that self-assemble in Winfree's abstract tile assembly model at temperature 1. We introduced the notion of a pumpable temperature 1 tile assembly system and then proved that, if $X$ weakly self-assembles in a pumpable, directed tile assembly system, then $X$ is necessarily "simple" in the sense that $X$ is merely a finite union of semi-doubly periodic sets. Finally, we conjecture that our results hold in the absence of the pumpability hypothesis.

**Conjecture 6.1.** *Let $\mathcal{T} = (T, \sigma, 1)$ be a directed tile assembly system and $\alpha \in \mathcal{A}_{\Box}[\mathcal{T}]$. If $\text{dom}\ \alpha$ is infinite, then $\mathcal{T}$ is pumpable.*

It is always possible to produce long paths in which the presence of a segment with two repetitions of a tile type does not imply that the segment is pumpable. However, in every case we consider, there is always a previous segment of the path that is pumpable. Proving Conjecture 6.1 would imply that every directed tile set weakly self-assembles a finite union of doubly periodic sets.

We also leave open the question of whether the hypothesis of directedness may be removed. We use the property of directedness at many points in our proof, but in some cases, a more careful and technically convoluted argument could be used to show that the tile set need not be directed.

Intuitively, an undirected tile set $\mathcal{T}$ that weakly self-assembles a set $X \subseteq \mathbb{Z}^2$ is deterministic in that all terminal assemblies of $\mathcal{T}$ "paint" exactly the points in $X$ black, but is nondeterministic in the sense that different terminal assemblies of $\mathcal{T}$ may place different tiles in the same location (including different black tiles at locations in $X$), and may even place non-black tiles at locations in one terminal assembly that are left empty in other terminal assemblies. Undirected tile assembly systems that weakly self-assemble a unique set $X$ exist, but in every case that we know of, the undirected tile set may be replaced by a directed tile set self-assembling the same set.

If both hypotheses of directedness and pumpability could be removed from the entire proof, then our main result would settle the case of computation via self-assembly at temperature 1, by showing that every temperature 1 tile assembly system weakly self-assembles a finite union of semi-doubly periodic sets if it weakly self-assembles any set at all. As indicated in the introduction, we interpret this statement to imply that general-purpose deterministic computation is not possible with temperature 1 tile assembly systems.

Finally, we make the observation that universal computation at temperature 1 *is* possible by changing the underlying model of self-assembly. For instance, a construction that was personally communicated by Matt Cook (also mentioned briefly in [4]) establishes that the abstract TAM is computationally universal with respect to directed, temperature 1 tile assembly systems that place tiles in three spatial dimensions. In fact, Cook's construction only uses two integer planes, "stacked" one on top of the other, to simulate an arbitrary cellular automaton (no such directed two-dimensional universality result is known at the time of this writing). Moreover, universality can also be achieved if negative glue strengths and interaction between differently colored glues (a so-called *non-diagonal strength function*) are allowed.

**Theorem 6.2.** *For every single-tape Turing machine $M$, there is a tile set $T$ with negative, non-diagonal glue strengths, which simulates $M$ in the following way. Given an input string $x$, define $\mathcal{T}_x = (T, \sigma_x, 1)$ to be the temperature 1 TAS where $\sigma_x$ is the seed assembly satisfying* $\mathrm{dom}\ \sigma_x = \{0, \ldots, |x|-1\} \times \{0\}$ *that encodes the initial configuration of $M$. Then $\mathcal{T}_x$ simulates the computation of $M(x)$, with the configuration of $M(x)$ after $n$ steps represented by the line $y = n$ in the terminal assembly of $\mathcal{T}_x$.*

A proof sketch of Theorem 6.2 is given in the technical appendix. Intuitively, by introducing negative glue strengths, we allow for the cooperation that is impossible with only nonnegative glue strengths. The difference with temperature 2 is that, at temperature 2, one may enforce that no tile appears at a certain position until two neighbors are present. At temperature 1 and with negative glue strengths, on the contrary, we do not enforce that two tiles are present before a position can be given a tile. However, we enforce that if the wrong tile binds in this position, eventually the error is corrected by the presence of a neighboring tile which forces the removal of the incorrect tile using negative glue strengths.

### Acknowledgment

We wish to thank Maria Axenovich, Matt Cook, and Jack Lutz for useful discussions. We would especially like to thank Niall Murphy, Turlough Neary and Damien Woods for inviting us to present a preliminary version of this research at the International Workshop on The Complexity of Simple Programs, University College Cork, Ireland on December 6th and 7th, 2008.

# References

[1] Leonard M. Adleman, Jarkko Kari, Lila Kari, and Dustin Reishus, *On the decidability of self-assembly of infinite ribbons*, Proceedings of the 43rd Annual IEEE Symposium on Foundations of Computer Science, 2002, pp. 530–537.

[2] James I. Lathrop, Jack H. Lutz, and Scott M. Summers, *Strict self-assembly of discrete Sierpinski triangles*, Theoretical Computer Science **410** (2009), 384–405.

[3] Matthew J. Patitz and Scott M. Summers, *Self-assembly of decidable sets*, Proceedings of The Seventh International Conference on Unconventional Computation (Vienna, Austria, August 25-28, 2008), 2008.

[4] Paul W. K. Rothemund, *Theory and experiments in algorithmic self-assembly*, Ph.D. thesis, University of Southern California, December 2001.

[5] Paul W. K. Rothemund and Erik Winfree, *The program-size complexity of self-assembled squares (extended abstract)*, STOC '00: Proceedings of the thirty-second annual ACM Symposium on Theory of Computing (New York, NY, USA), ACM, 2000, pp. 459–468.

[6] Paul W.K. Rothemund, Nick Papadakis, and Erik Winfree, *Algorithmic self-assembly of dna sierpinski triangles*, PLoS Biology **2** (2004), no. 12.

[7] Nadrian C. Seeman, *Nucleic-acid junctions and lattices*, Journal of Theoretical Biology **99** (1982), 237–247.

[8] Hao Wang, *Proving theorems by pattern recognition – II*, The Bell System Technical Journal **XL** (1961), no. 1, 1–41.

[9] ———, *Dominoes and the AEA case of the decision problem*, Proceedings of the Symposium on Mathematical Theory of Automata (New York, 1962), Polytechnic Press of Polytechnic Inst. of Brooklyn, Brooklyn, N.Y., 1963, pp. 23–55.

[10] Erik Winfree, *Algorithmic self-assembly of DNA*, Ph.D. thesis, California Institute of Technology, June 1998.

# 7 Technical Appendix

This appendix contains proofs of results that were not given in the main text, and some definitions required only for those proofs.

*Proof.* (*of Observation 4.2*) Let $B \subseteq \mathbb{Z}^2$ be a semi-doubly periodic set. It is routine to verify that $B$'s projection along the $x$-axis is a (singly) periodic set; i.e., a set

$$B_x = \{\ x \in \mathbb{Z} \mid (x, y) \in B \text{ for some } y \in \mathbb{Z}\ \}$$

such that there is a number $v \in \mathbb{N}$ such that, for all $x \in \mathbb{Z}$, $x \in B_x \implies x+v \in B_x$. It is well-known that a unary language $L \in \{0\}^*$ is regular if and only if the set $N = \{\ n \in \mathbb{N} \mid 0^n \in L\ \}$ of lengths of strings in $L$ is eventually periodic. Therefore the language $L_{B,x} = \left\{\ 0^{|n|}\ \middle|\ n \in B_x\ \right\}$ is regular. A symmetric argument establishes that $L_{B,y}$ is a regular language as well, so the theorem holds for any semi-doubly periodic set. Since $A$ is a finite union of semi-doubly periodic sets, the theorem follows by the closure of the regular languages under finite union. □

We will now prove a series of technical lemmas that will reveal "order" in the seemingly disordered realm of temperature 1 (a.k.a., non-cooperative) self-assembly.

**Definition 7.1.** Let $\alpha_1, \alpha_2$ be assemblies. We say that $\alpha_1$ and $\alpha_2$ are *consistent* if, for all $\vec{v} \in \text{dom}\ \alpha_1 \cap \text{dom}\ \alpha_2$, $\alpha_1(\vec{v}) = \alpha_2(\vec{v})$.

**Definition 7.2.** Let $\alpha_1, \alpha_2$ be consistent assemblies. The *union of* $\alpha_1$ *and* $\alpha_2$, written as $\alpha_1 \cup \alpha_2$, is the unique assembly $\alpha_1 \cup \alpha_2$ satisfying, $\text{dom}\ (\alpha_1 \cup \alpha_2) = \text{dom}\ \alpha_1 \cup \text{dom}\ \alpha_2$ and $\alpha_1 \sqsubseteq \alpha_1 \cup \alpha_2$ and $\alpha_2 \sqsubseteq \alpha_1 \cup \alpha_2$.

**Observation 7.3.** *Let $\mathcal{T} = (T, \sigma, 1)$ be a directed TAS. If $\alpha_1$ and $\alpha_2$ are consistent, connected assemblies, and $\alpha_1 \in \mathcal{A}[\mathcal{T}]$, then $\alpha_1 \cup \alpha_2 \in \mathcal{A}[\mathcal{T}]$.*

The next lemma states that, if we wish to show that the existence of a semi-periodic path in the assembly implies a periodic path (i.e., it is not only one-way infinite but two-way infinite, bisecting the plane), it suffices to show that a translation – along the path – of the assembly connecting the seed to the path, are consistent with the path, so long as the translation is sufficiently far to achieve maximal intersection between the assembly and the path.

**Lemma 7.4.** *Let $\mathcal{T} = (T, \sigma, 1)$ be a directed TAS, $\vec{0} \neq \vec{v} \in \mathbb{Z}^2$, $\alpha$ be the unique assembly satisfying $\alpha \in \mathcal{A}_\Box[\mathcal{T}]$, and $\vec{a} \in \text{dom}\ \alpha$. Let $\pi_{\vec{a}}^{\rightarrow}$ be a $\vec{v}$-periodic path in $\alpha$ originating at $\vec{a}$, and let $\pi_{\vec{0},\vec{a}}$ be a simple finite path in $G_\alpha$ from $\vec{0}$ to $\vec{a}$. Let $c > |\pi_{\vec{0},\vec{a}}|$ be a positive integer. If $\left(\alpha \upharpoonright \pi_{\vec{0},\vec{a}}\right) + c \cdot \vec{v}$ is consistent with $\alpha \upharpoonright \pi_{\vec{a}}^{\rightarrow}$, then there is a (two-way infinite) $\vec{v}$-periodic path in $G_\alpha$ containing $\vec{a}$.*

*Proof.* Since $c > \left|\pi_{\vec{0},\vec{a}}\right|$, $\text{dom}\left(\left(\alpha \upharpoonright \pi_{\vec{0},\vec{a}}\right) + c \cdot \vec{v}\right) \cap \pi_{\vec{a}}^{\rightarrow}$ is maximal over all $c \in \mathbb{N}$. Since $\left(\alpha \upharpoonright \pi_{\vec{0},\vec{a}}\right) + c \cdot \vec{v}$ is consistent with $\alpha \upharpoonright \pi_{\vec{a}}^{\rightarrow}$, extending the tiles of $\pi_{\vec{a}}^{\rightarrow}$ in the direction $-\vec{v}$ will intersect $\pi_{\vec{0},\vec{a}}$ only at each position where the existing tile in $\alpha \upharpoonright \pi_{\vec{0},\vec{a}}$ agrees with the extension. Therefore such an extension will not be blocked by $\alpha \upharpoonright \pi_{\vec{0},\vec{a}}$, and $\pi_{\vec{a}}^{\rightarrow}$ is merely one half of a (two-way) periodic path in $\alpha$. □

The next lemma states that if two parallel semi-periodic paths are connected through their pumpable segments in "another way besides the trivial way" (since all points are connected by *some* path in the binding graph), then this connection can be exploited to show that the paths must actually form two-way periodic paths that bisect the plane.

**Lemma 7.5.** *Let $\mathcal{T} = (T, \sigma, 1)$ be a directed TAS, $\vec{0} \neq \vec{v} \in \mathbb{Z}^2$, $\alpha$ be the unique assembly satisfying $\alpha \in \mathcal{A}_{\Box}[\mathcal{T}]$, $\vec{a} \in \mathrm{dom}\ \alpha$ and $\pi_{\vec{0},\vec{a}}$ be a simple finite path from $\vec{0}$ to $\vec{a}$ in $\alpha$. If $\pi_{\vec{a}}^{\rightarrow}$ is a $\vec{v}$-semi-periodic path in $G_\alpha$ originating at $\vec{a}$, $\vec{d} \in \mathrm{dom}\ \alpha$ such that $\pi_{\vec{d}}^{\rightarrow}$ is a $z \cdot \vec{v}$-semi-periodic path in $\alpha$ originating at $\vec{d}$, for some $z \in \mathbb{R}^+$, and there is a simple path from some point $\vec{c}$ on the tail of $\pi_{\vec{a}}^{\rightarrow}$ to $\vec{d}$ in $G_\alpha$, denoted as $\pi_{\vec{c},\vec{d}}$, with $\pi_{\vec{c},\vec{d}} \cap \left(\pi_{\vec{0},\vec{a}} \cup \pi_{\vec{a}}^{\rightarrow} \cup \pi_{\vec{d}}^{\rightarrow}\right) = \left\{\vec{c}, \vec{d}\right\}$, then there exists a $\vec{v}$-periodic path in $\alpha$ containing $\vec{a}$.*

Intuitively, we prove Lemma 7.5 by showing that an appropriately translated copy of the simple (finite) path from the seed to the origination point of $\pi_{\vec{a}}^{\rightarrow}$ is consistent with $\pi_{\vec{a}}^{\rightarrow}$. See Figure 5 for a visual depiction of (a simple example of) the hypothesis of Lemma 7.5 and Figure 6 for an illustration of the conclusion of Lemma 7.5.

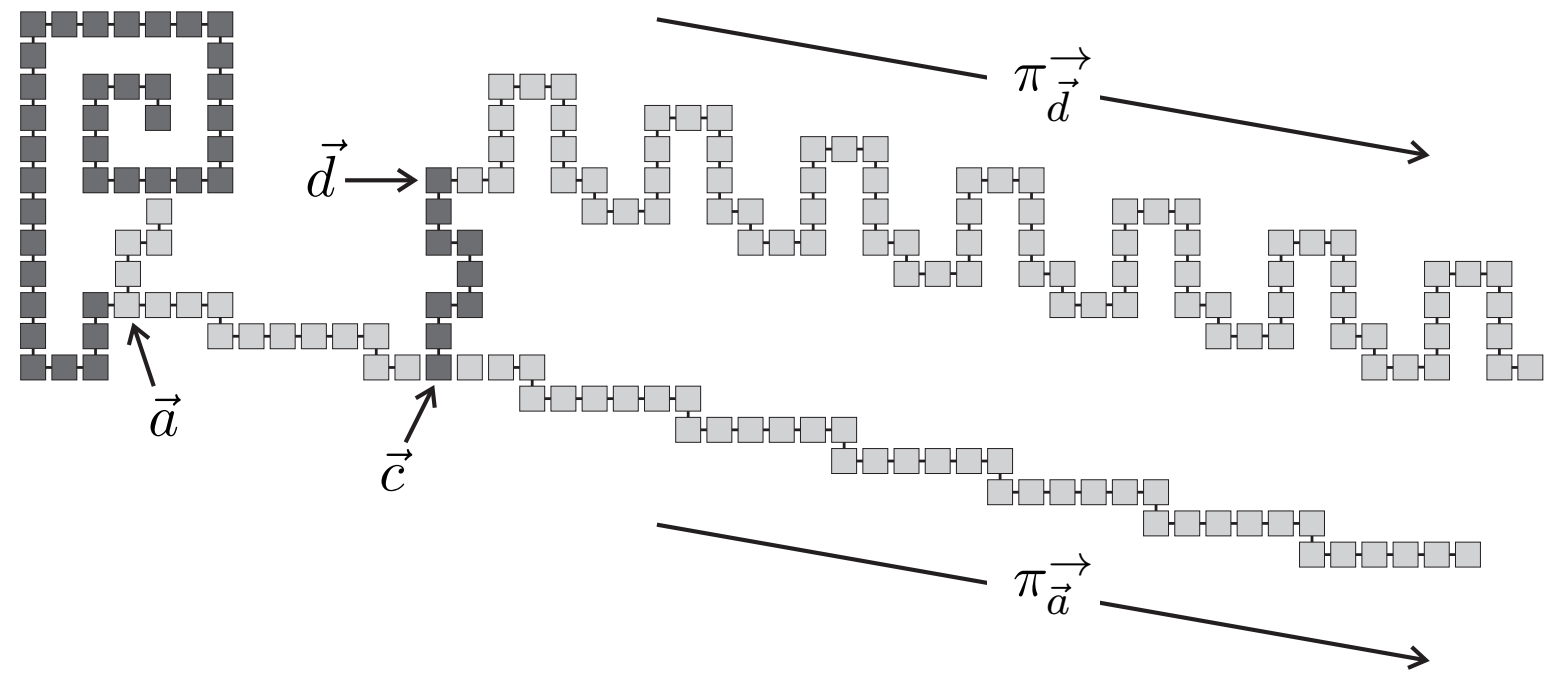

Figure 5: An example of the hypothesis of Lemma 7.5. The seed tile is at the center of the "spiral." The path $\pi_{\vec{c},\vec{d}}$ is represented by the dark tiles that form the "bridge" between the tail of $\pi_{\vec{a}}^{\rightarrow}$ (the lowest $\vec{v}$-periodic path) and some point on $\pi_{\vec{d}}^{\rightarrow}$ (the upper $\vec{v}$-periodic path).

*Proof.* Let $c \in \mathbb{N}$ such that $c > \left|\pi_{\vec{0},\vec{a}}\right|$. By Lemma 7.4, it suffices to show that $\left(\alpha \upharpoonright \pi_{\vec{0},\vec{a}}\right) + c \cdot \vec{v}$ is consistent with $\alpha \upharpoonright \pi_{\vec{a}}^{\rightarrow}$. If consistency holds, then we are done, so assume otherwise. This means that there is a point

$$\vec{b} \in \left(\pi_{\vec{0},\vec{a}} + c \cdot \vec{v}\right) \cap \pi_{\vec{a}}^{\rightarrow},$$

satisfying the following two conditions.

1. $\left(\left(\alpha \upharpoonright \pi_{\vec{0},\vec{a}}\right) + c \cdot \vec{v}\right)\left(\vec{b}\right) \neq \left(\alpha \upharpoonright \pi_{\vec{a}}^{\rightarrow}\right)\left(\vec{b}\right)$, and

2. $\vec{b}$ is the "closest" point to $\vec{a} + c \cdot \vec{v}$ on the path $\pi_{\vec{0},\vec{a}} + c \cdot \vec{v}$.

Moreover, the assumption that $\pi_{\vec{c},\vec{d}} \cap \left(\pi_{\vec{0},\vec{a}} \cup \pi_{\vec{a}}^{\rightarrow} \cup \pi'\right) = \left\{\vec{c}, \vec{d}\right\}$, along with the existence of the $\vec{v}$-periodic path $\pi_{\vec{d}}^{\rightarrow}$, can be used to show that there exists a simple cycle $C$ in $G_\alpha$ that contains the point $\vec{b}$ such that not every simple (finite) path from $\vec{0}$ to (any point in) $C$ goes through $\vec{b}$. This means that it is possible to define an assembly sequence where (the tile placed at) the input neighbor of $\vec{b}$ is the same as the (tile placed at) the output neighbor of $\vec{b} - c \cdot \vec{v}$ in the assembly sequence resulting in $\alpha \upharpoonright \pi_{\vec{0},\vec{a}}$. Since $\mathcal{T}$ is directed, these must be the same tiles, whence there can be no "first" point of inconsistency between $\left(\alpha \upharpoonright \pi_{\vec{0},\vec{a}}\right) + c \cdot \vec{v}$ and $\alpha \upharpoonright \pi_{\vec{a}}^{\rightarrow}$. $\square$

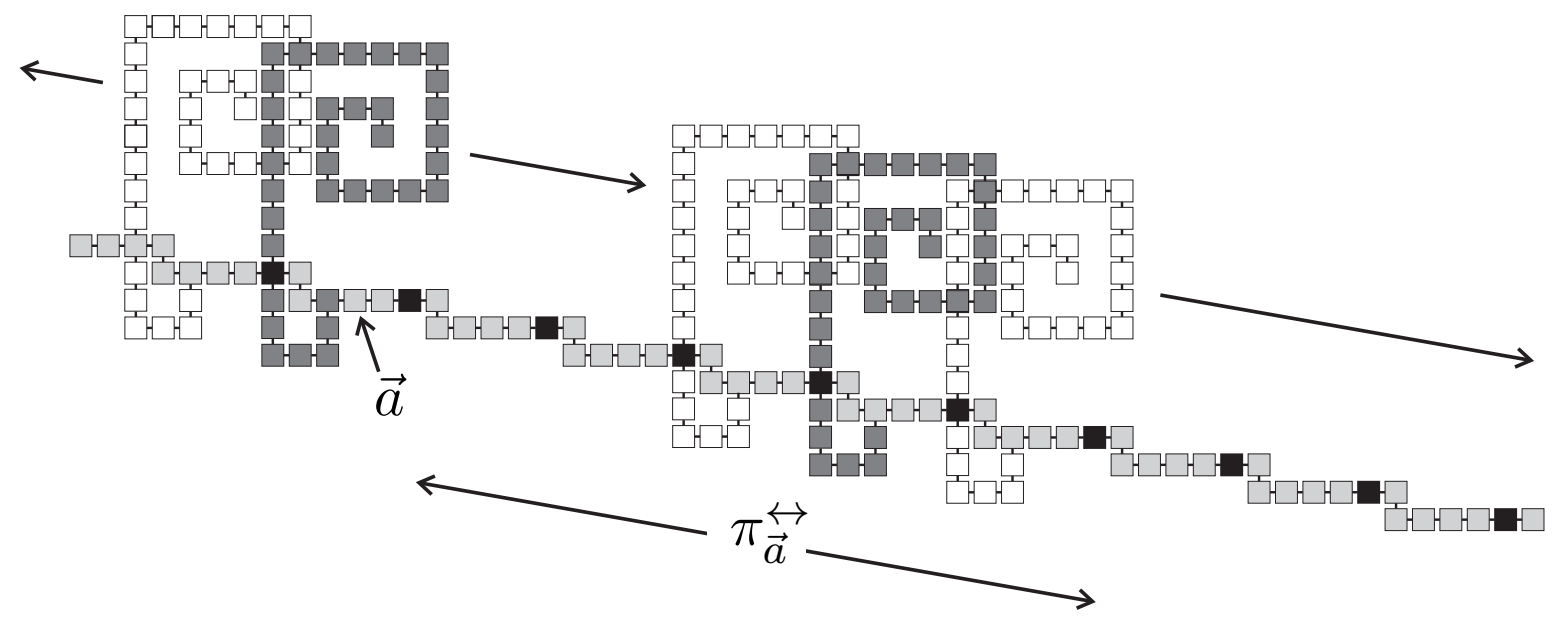

Figure 6: An example of the conclusion of Lemma 7.5. Note that the path $\pi_{\vec{a}}^{\rightarrow}$ is a (two-way infinite) $\vec{v}$-periodic path, which means that (copies of) the path $\pi_{\vec{0},\vec{a}}$ can be translated in the $\pm\vec{v}$ direction.

**Definition 7.6.** Let $\alpha$ be an assembly. We say $\alpha$ is *doubly periodic* if there exist $\vec{u}, \vec{v} \in \mathbb{Z}^2$ such that $\vec{u} \neq \vec{v}$, $\vec{u} \neq \vec{0}$, $\vec{v} \neq \vec{0}$, and for all $\vec{a} \in \mathbb{Z}^2$, $\alpha(\vec{a}) = \alpha(\vec{a} + \vec{u}) = \alpha(\vec{a} + \vec{v})$, where $\alpha(\vec{a}) = \alpha(\vec{b})$ if both $\alpha(\vec{a})$ and $\alpha(\vec{b})$ are both undefined.

In other words, $\alpha$ is doubly periodic if its values form a repeating "grid" of parallelogram patterns on $\mathbb{Z}^2$, infinite in all directions. It is easy to see that, if $\alpha$ is doubly periodic, then for any set of tile types $B \subseteq T$, dom $\alpha(B)$ (the set that weakly self-assembles) is a union of four semi-doubly periodic sets of integer lattice points.

The following technical lemma shows that, if a tail of a tooth of a comb connects to the base through a path other than the "natural" one, then the entire assembly is doubly periodic. This allows us to assume, in the proof of our main theorem, that the teeth of a comb are not interconnected in "inconvenient" ways.

**Lemma 7.7.** *Let $\mathcal{T} = (T, \sigma, 1)$ be a directed TAS, $\alpha \in \mathcal{A}_{\Box}[\mathcal{T}]$, and $\alpha^*$ be a comb in $\alpha$. If the tail of the $n^{th}$ tooth is connected to (any point in) the $(n+1)^{st}$ tooth (for $n > 1$) via a simple finite path that does not go through any point on the $n^{th}$ tooth, then $\alpha$ is doubly periodic.*

The proof idea of Lemma 7.7 is as follows. Suppose that $\alpha$ is the unique terminal assembly satisfying $\alpha \in \mathcal{A}_{\Box}[\mathcal{T}]$. Since the tails of two different teeth of the comb $\alpha^*$ are connected via a simple path, we can build an assembly sequence whose result, denoted as $\alpha_{\#}$, is an infinite, plane-filling, doubly periodic assembly. We then show that this assembly is consistent with the tiles near the seed, which means it must be a producible assembly. Finally, since $\mathcal{T}$ is directed and $\alpha_{\#} \sqsubseteq \alpha$, we conclude that $\alpha$ is doubly periodic. Figure 7 illustrates (a simple example of) the hypothesis of the lemma.

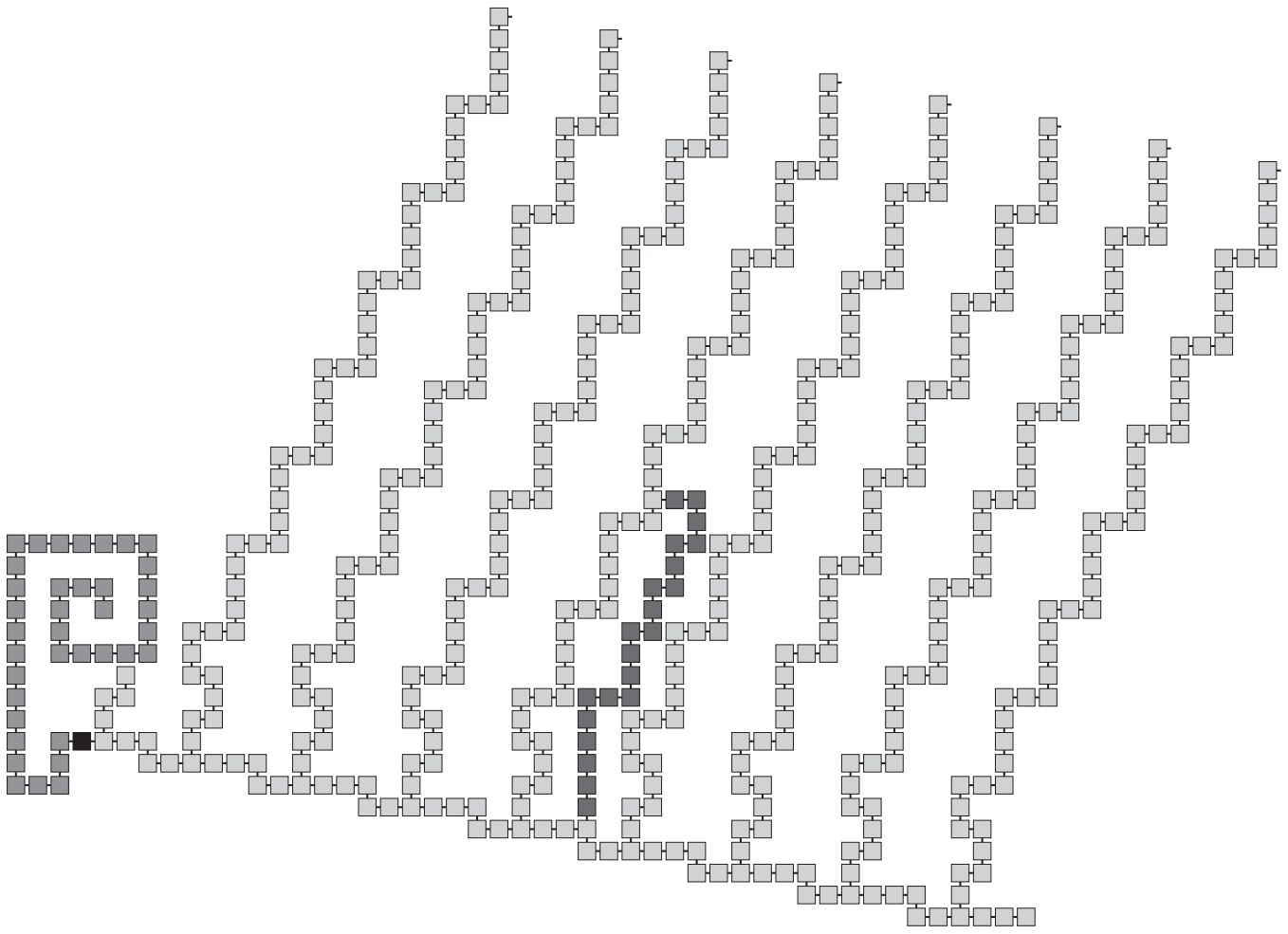

Figure 7: An example of a comb having two teeth connected via a simple path that does not go through the finite closure of the base. The seed tile is the tile in the center of the "spiral".

*Proof.* Let $\alpha^*$ be a comb in $\alpha$ with respect to $\pi^{\rightarrow}$ and $\pi^{\uparrow}$, and $\vec{a}$ be the starting point of $\alpha^*$. Assume that, for non-zero vectors $\vec{u} \neq \vec{v}$, $\pi^{\rightarrow}$ is an eventually $\vec{u}$-semi-periodic path in $\alpha$ originating at $\vec{0}$, and $\pi^{\uparrow}$ is an eventually $\vec{v}$-semi-periodic path in $\alpha$ originating at $\vec{b} \in \pi^{\rightarrow}$ with $\pi^{\rightarrow} \cap \pi^{\uparrow} = \left\{\vec{b}\right\}$.

The hypothesis says that, for some $1 < n \in \mathbb{N}$, there is a simple finite path from some point $\vec{c}$ on the tail of $\pi^{\uparrow} + n \cdot \vec{u}$ to some point $\vec{d} \in \pi^{\uparrow} + (n+1) \cdot \vec{u}$ that does not go through any point in $\pi^{\uparrow} + n \cdot \vec{u}$ except for the first point on $\pi_{\vec{c},\vec{d}}$. Denote this path as $\pi_{\vec{c},\vec{d}}$. Lemma 7.5 tells us that there exists a (two-way infinite) $\vec{v}$-periodic path in $\alpha$ containing the point $\widehat{\vec{b}} = \vec{b} + n \cdot \vec{u}$. Denote this path as $\pi^{\updownarrow}_{\widehat{\vec{b}}}$.

Let $\pi^{\rightarrow}_{\widehat{\vec{b}}}$ be the $\vec{u}$-semi-periodic path in $G_\alpha$ originating at $\widehat{\vec{b}}$. Since $\pi^{\updownarrow}_{\widehat{\vec{b}}}$ is $\vec{v}$-periodic and $\pi^{\updownarrow}_{\widehat{\vec{b}}} \cap \pi^{\rightarrow} \neq \varnothing$, it must be the case that there is a $\vec{u}$-semi-periodic path, denoted as $\pi^{\rightarrow}_{\widehat{\vec{b}}+\vec{v}}$, in $\alpha$ originating at the point $\widehat{\vec{b}} + \vec{v}$. Let $\pi_{\vec{0},\widehat{\vec{b}}}$ be a finite simple path from $\vec{0}$ to $\widehat{\vec{b}}$ in $G_\alpha$. If we (re)define $\vec{c} = \widehat{\vec{b}}$, $\vec{d} = \widehat{\vec{b}} + \vec{v}$, and note that there is a simple (finite) path from $\vec{c}$ on the tail of $\pi^{\rightarrow}$ (because $n > 1$) to $\vec{d} \in \pi^{\rightarrow}_{\widehat{\vec{b}}+\vec{v}}$, denoted as $\pi_{\vec{c},\vec{d}}$, in $G_\alpha$ with $\pi_{\vec{c},\vec{d}} \cap \left(\pi_{\vec{0},\widehat{\vec{b}}} \cup \pi^{\rightarrow}_{\widehat{\vec{b}}} \cup \pi^{\rightarrow}_{\widehat{\vec{b}}+\vec{v}}\right) = \left\{\vec{c}, \vec{d}\right\}$, then we can again use Lemma 7.5 to conclude that there is a (two-way infinite) $\vec{v}$-periodic path containing $\widehat{\vec{b}}$, hence also containing $\vec{a}$ since those points differ by a multiple of $\vec{v}$. Denote this path as $\pi^{\leftrightarrow}_{\vec{a}}$. If $\pi_{\vec{0},\vec{a}}$ is the initial segment of $\pi^{\rightarrow}$, then the assembly $\alpha \upharpoonright \left(\pi_{\vec{0},\vec{a}} \cup \pi^{\leftrightarrow}_{\vec{a}} \cup \pi^{\updownarrow}_{\vec{b}}\right) \in \mathcal{A}[\mathcal{T}]$.

Define the assembly $\alpha_{\#}$ as follows.

$$\text{dom } \alpha_{\#} = \bigcup_{n \in \mathbb{Z}} \left((\alpha \upharpoonright \pi^{\leftrightarrow}_{\vec{a}}) + n \cdot \vec{v}\right) \cup \bigcup_{n \in \mathbb{Z}} \left(\left(\alpha \upharpoonright \pi^{\updownarrow}_{\vec{b}}\right) + n \cdot \vec{u}\right).$$

Our goal is to show that $\alpha_{\#} \cup \left(\alpha \upharpoonright \pi_{\vec{0},\vec{a}}\right) \in \mathcal{A}[\mathcal{T}]$, which will nearly complete the proof since $\alpha_{\#}$ is doubly periodic.

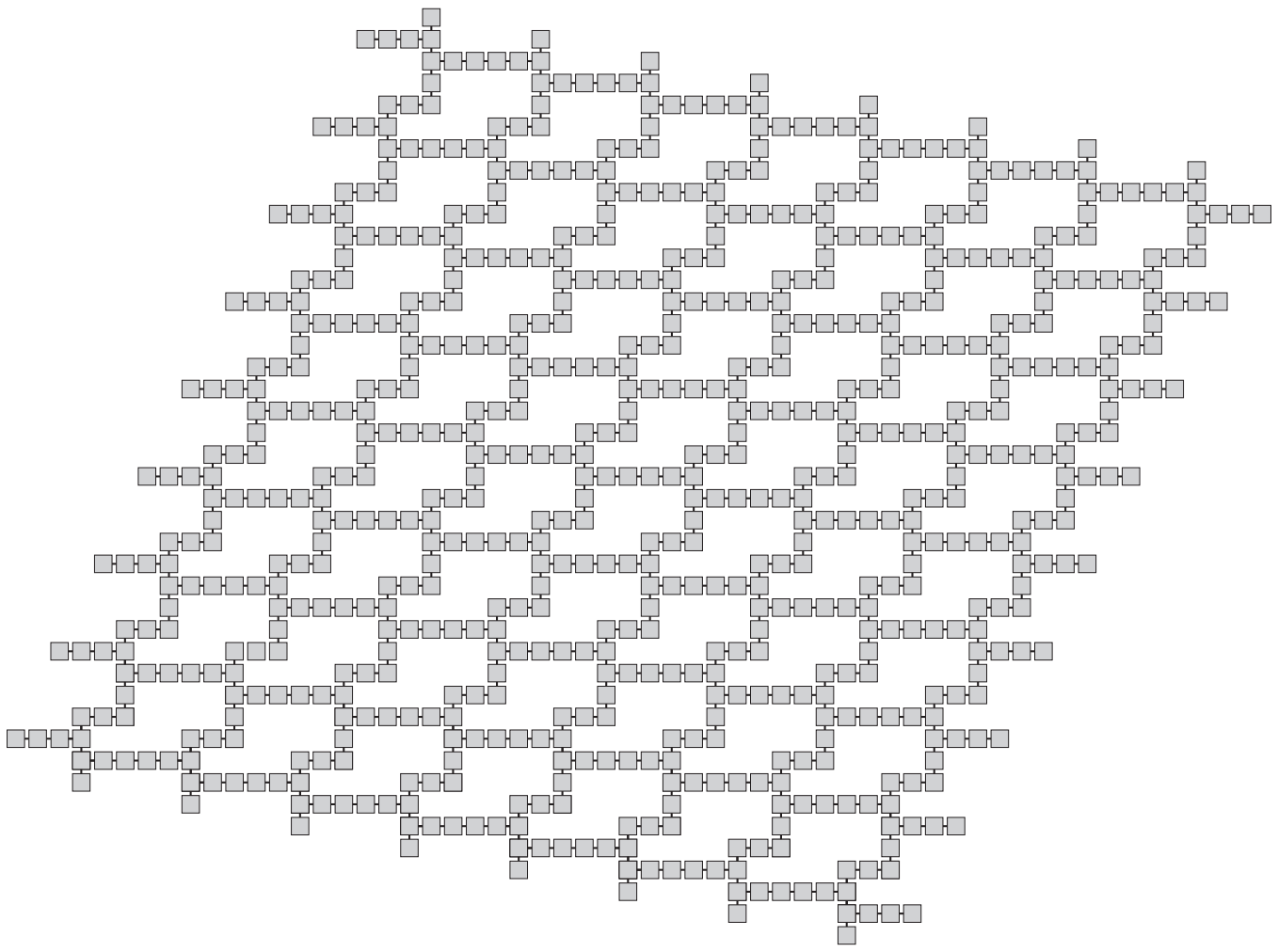

Figure 8: A finite sub-assembly of $\alpha_{\#}$.

Let $\alpha_{\text{almost-}\#}$ be the largest assembly satisfying $\alpha_{\text{almost-}\#} \sqsubseteq \alpha_{\#}$ such that $\text{dom } \alpha_{\text{almost-}\#} \cap \pi_{\vec{0},\vec{a}} = \varnothing$ and $G_{\alpha_{\text{almost-}\#}}$ is connected. Note that the assembly $\alpha_{\text{almost-}\#} \cup \left(\alpha \restriction \pi_{\vec{0},\vec{a}}\right) \in \mathcal{A}[\mathcal{T}]$. See Figure 9 for an example of how the assembly $\alpha_{\text{almost-}\#}$ is constructed.

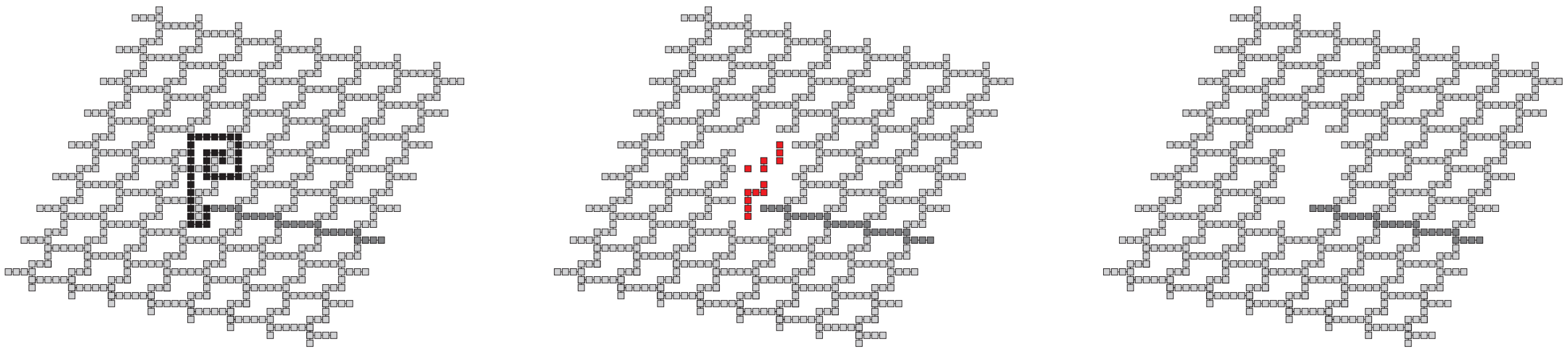

(a) $\alpha_{\#} \cup \left(\alpha \restriction \pi_{\vec{0},\vec{a}}\right)$ (b) $\alpha_{\#} - \left(\alpha \restriction \pi_{\vec{0},\vec{a}}\right)$ (c) $\alpha_{\text{almost}-\#}$

Figure 9: The black tiles are the path $\pi_{\vec{0},\vec{a}}$. Since $\alpha_{\#} - \left(\alpha \restriction \pi_{\vec{0},\vec{a}}\right)$ may have a finite number of disconnected "islands", we remove these as well to get the largest *connected* subassembly of $\alpha_{\#}$.

In order to prove that $\alpha_{\#} \cup \left(\alpha \restriction \pi_{\vec{0},\vec{a}}\right) \in \mathcal{A}[\mathcal{T}]$, it suffices to show that $\alpha \restriction \pi_{\vec{0},\vec{a}}$ and $\alpha_{\#}$ are consistent, since $\alpha \restriction \pi_{\vec{0},\vec{a}} \in \mathcal{A}[\mathcal{T}]$ and $\pi_{\vec{0},\vec{a}} \cap \text{dom } \alpha_{\#} \neq \varnothing$. Note that, by the way we constructed $\alpha_{\#}$, not *every* point in $\pi_{\vec{0},\vec{a}}$ can be a point of inconsistency between $\alpha_{\#}$ and $\alpha \restriction \pi_{\vec{0},\vec{a}}$. Fix $n^* \in \mathbb{N}$ such that $\pi_{\vec{0},\vec{a}} \cap \left(\pi_{\vec{0},\vec{a}} + (n^* \cdot \vec{u} + n^* \cdot \vec{v})\right) = \varnothing$, which exists by the finiteness of $\pi_{\vec{0},\vec{a}}$. Since there exists at least one point $\vec{s} \in \pi_{\vec{0},\vec{a}} \cap \text{dom } \alpha_{\#}$ such that $\alpha_{\#}(\vec{s}) = \left(\alpha \restriction \pi_{\vec{0},\vec{a}}\right)(\vec{s})$, we can conclude that

$$\left(\alpha_{\text{almost-}\#} \cup \left(\alpha \restriction \pi_{\vec{0},\vec{a}}\right)\right) \cup \left(\left(\alpha \restriction \pi_{\vec{0},\vec{a}}\right) + n^* \cdot \vec{u} + n^* \cdot \vec{v}\right) \in \mathcal{A}[\mathcal{T}]$$

because $\mathcal{T}$ is directed and every point in $\pi_{\vec{0},\vec{a}} + n^* \cdot \vec{u} + n^* \cdot \vec{v} \cap \text{dom } \alpha_{\text{almost-}\#}$ is reachable from $\vec{0}$ via at least two simple finite paths. Thus, $\alpha_{\#}$ and $\alpha \restriction \pi_{\vec{0},\vec{a}}$ must be consistent. Since $\alpha_{\#}$ is trivially doubly periodic, it follows that $\alpha_{\#} \cup \left(\alpha \restriction \pi_{\vec{0},\vec{a}}\right) \in \mathcal{A}[\mathcal{T}]$.

Since $\alpha_{\#} \sqsubseteq \alpha$, it suffices to show that the tiles added to $\alpha_{\#}$ to create $\alpha$ are doubly periodic. Since $\alpha_{\#}$ forms an infinite grid that partitions $\mathbb{Z}^2$ into infinitely many identical "parallelograms", bordered by the same tiles $\left(\text{those forming a single period of } \pi^{\leftrightarrow}_{\vec{a}} \text{ and } \pi^{\updownarrow}_{\vec{b}}, \text{ respectively}\right)$, then by the directedness of $\mathcal{T}$, each of these parallelograms must have the same tiles in the same positions relative to the borders of the parallelogram. Since the parallelogram borders are doubly periodic, the contents within the borders are doubly periodic as well, whence $\alpha$ is doubly periodic. □

The following lemma states that if there is a path from an assembly $\alpha'$ to a point not on the finite closure of $\alpha'$, then there is an eventually periodic path, originating from the same point as the first path, that intersects $\alpha'$ only at the first point.

**Lemma 7.8.** *Let $\mathcal{T} = (T, \sigma, 1)$ be a pumpable directed TAS, $\alpha$ be the unique assembly satisfying $\alpha \in \mathcal{A}_{\Box}[\mathcal{T}]$, $\alpha' \sqsubseteq \alpha$, and $\vec{x} \in \text{dom } \alpha$. If $\vec{x} \notin \mathcal{F}(\alpha')$ and there is a finite simple path in $G_\alpha$ that goes through some point $\vec{s} \in \alpha'$ to $\vec{x}$, then, for some $\vec{0} \neq \vec{v} \in \mathbb{Z}^2$, there exists an eventually $\vec{v}$-semi-periodic path $\pi$ in $G_\alpha$, originating at $\vec{s}$, with $\pi \cap \text{dom } \alpha' = \{\vec{s}\}$.*

*Proof.* Let $\vec{r} \in \text{dom } \alpha'$ be a point connected to $\vec{x}$ by a simple finite path $\pi_{\vec{r},\vec{x}}$ satisfying $\pi_{\vec{r},\vec{x}} \cap \text{dom } \alpha' = \{\vec{r}\}$. Let $\vec{p}_0$ be the point on $\pi_{\vec{r},\vec{x}}$ closest to $\vec{x}$ such that $G_\alpha$ has an infinite simple path $\pi_{\vec{p}_0,\infty}$ starting at $\vec{p}_0$ that does not contain the point immediately before $\vec{p}_0$ on $\pi_{\vec{r},\vec{x}}$. Let $\vec{p}_1 \in \text{dom } \alpha \cap \left(\pi_{\vec{p}_0,\infty} - D(c, \vec{p}_0)\right)$, and $\pi_{\vec{r},\vec{p}_0,\vec{p}_1}$ be a simple path in $G_\alpha$ from $\vec{r}$ to $\vec{p}_1$ that goes through $\vec{p}_0$. Since $\mathcal{T}$ is pumpable, and $\left(\pi_{\vec{p}_0,\infty} - D(c, \vec{p}_0)\right)$, $\pi_{\vec{r},\vec{p}_0,\vec{p}_1}$ is a pumpable path. Let $\pi_{\vec{r},\vec{p}_0,\vec{p}_1}[i \ldots k]$, for $0 \leq i < k \leq \left|\pi_{\vec{r},\vec{p}_0,\vec{p}_1}\right|$, be the first pumpable segment of $\pi_{\vec{r},\vec{p}_0,\vec{p}_1}$, and $\vec{v} = \pi_{\vec{r},\vec{p}_0,\vec{p}_1}[k] - \pi_{\vec{r},\vec{p}_0,\vec{p}_1}[i]$. The lemma follows by letting $\vec{s} = \pi_{\vec{r},\vec{p}_0,\vec{p}_1}[i]$, and $\pi$ be the $\vec{v}$-semi-periodic path originating at the point $\vec{s}$ defined by the pumpable segment. □

The next lemma states that if two teeth in a comb are connected in an "inconvenient" way (i.e., other than the trivial ways in which any two points in a stable assembly must be connected), then the entire assembly is an (two-way, two-dimensional) doubly periodic grid.

**Lemma 7.9.** *Let $\mathcal{T} = (T, \sigma, 1)$ be a pumpable directed TAS, $\alpha$ be the unique assembly satisfying $\alpha \in \mathcal{A}_{\Box}[\mathcal{T}]$, and $\alpha^*$ be a comb in $\alpha$. If there exists $\vec{x} \in \text{dom } \alpha$ such that there is a simple finite path from $\vec{0}$ to $\vec{x}$ that goes through the tail of some tooth of $\alpha^*$, then $\vec{x} \in \mathcal{F}(\alpha^*)$ or $\alpha$ is doubly periodic.*

*Proof.* Let $\pi^{\uparrow}$ be the first tooth of $\alpha^*$. Assume that $\vec{x} \notin \mathcal{F}(\alpha^*)$. It suffices to show that $\alpha$ is doubly periodic. Let $\vec{r}$ be an element of the tail of $n \cdot \vec{u} + \pi^{\uparrow}$ for some $n \in \mathbb{N}$, such that there exists a simple path in $G_\alpha$ from $\vec{r}$ to $\vec{x}$, denoted as $\pi_{\vec{r},\vec{x}}$, with $\pi_{\vec{r},\vec{x}} \cap \text{dom } \alpha^* = \{\vec{r}\}$. Such a point $\vec{r}$ exists because $\vec{x} \notin \mathcal{F}(\alpha^*)$. By Lemma 7.8, there exists an eventually $\vec{w}$-semi-periodic $\pi$ in $G_\alpha$, originating at $\vec{r}$ with $\pi \cap \text{dom } \alpha^* = \varnothing$.

If $\vec{w} \neq z \cdot \vec{v}$ for some $z \in \mathbb{R}^+$, then the $\vec{w}$-semi-periodic path it defines intersects a tooth of $\alpha^*$ and, by Lemma 7.7, $\alpha$ is doubly-periodic. Therefore assume that $\vec{w} = z \cdot \vec{v}$ for some $z \in \mathbb{R}^+$. Let $\vec{a} = \pi[s]$ - the point at which $\pi$ becomes periodic. Lemma 7.5 tells us that there exists a (two-way infinite) $\vec{w}$-periodic path in $\alpha$ containing the point $\vec{a}$. Denote this path as $\pi^{\updownarrow}_{\vec{a}}$. Let $\vec{b}$ be the closest point to $\vec{a}$ on the path $\pi^{\updownarrow}_{\vec{a}}$ that intersects the base of $\alpha^*$. Let $\widehat{\pi}^{\uparrow}$ be the $\vec{w}$-semi-periodic path originating at $\vec{b}$ that does not intersect any teeth of $\alpha^*$. Then we can form a new comb $\widehat{\alpha^*}$ with respect to $\pi^{\rightarrow}$ and $\widehat{\pi}^{\uparrow}$. Finally, we can apply Lemma 7.7 to either $\alpha^*$ or $\widehat{\alpha^*}$ in order to conclude that $\alpha$ is doubly periodic. □

The following observation is helpful in the proof of our main theorem, and states the obvious fact that combs define semi-doubly periodic sets.

**Observation 7.10.** *Let $\mathcal{T} = (T, \sigma, 1)$ be a directed pumpable TAS in which the set $X \subseteq \mathbb{Z}^2$ weakly self-assembles, and let $\alpha$ be the unique terminal assembly satisfying $\alpha \in \mathcal{A}_{\Box}[\mathcal{T}]$. If $\alpha^* \in \mathcal{A}[\mathcal{T}]$ is a comb in $\alpha$, then the set $X \cap \operatorname{dom} \alpha^*$ is a finite union of semi-doubly periodic sets.*

*Proof.* (*of Theorem 4.3*) Assume the hypothesis. Let $c \in \mathbb{N}$ testify to the $c$-pumpability of $\mathcal{T}$. We denote by $D_1$, $D_2$, and $D_3$ the diamonds of radius $c$, $2c$, and $3c$, respectively, around the origin.

We show that each $\vec{x} \in X$ outside of $D_2$ is part of the finite closure of some periodic path or comb originating in $D_3$. Therefore, the union of the black tiles of the finite closure of each comb starting in $D_3$, each periodic path starting in $D_3$, and each singleton set containing a black tile in $D_3$, is the union proving the theorem. There are only a finite number of points in $D_3$, though each could be the origination point of multiple combs or periodic paths. However, if any such point has an infinite number of combs or periodic paths originating from it, then some will intersect at a different point, creating a cycle between the tails of two teeth of combs, and Lemma 7.7 tells us that $X$ is doubly periodic. Otherwise, only a finite number of combs and periodic paths can originate in $D_3$, each one defining a term in the union, showing that the union is finite. Recall that a singleton set and a semi-periodic path are semi-doubly periodic sets, and Observation 7.10 allows us to conclude that the remaining terms representing combs are semi-doubly periodic as well.

Let $\vec{x} \in X - D_1$, and $\alpha$ be the unique assembly satisfying $\alpha \in \mathcal{A}_{\Box}[\mathcal{T}]$. Since $\vec{x} \notin D_1$, and because $\mathcal{T}$ is $c$-pumpable, there exists a $c$-pumpable path from $\vec{0}$ to $\vec{x}$ in $\alpha$. Denote this path as $\pi_{\vec{0},\vec{x}}$. Let $\pi_{\vec{0},\vec{x}}[i \dots k]$, for $0 \leq i < k \leq \left|\pi_{\vec{0},\vec{x}}\right|$, be the first pumpable segment of $\pi_{\vec{0},\vec{x}}$, and $\vec{u} = \pi_{\vec{0},\vec{x}}[k] - \pi_{\vec{0},\vec{x}}[i]$. Let $\pi^{\rightarrow}_{\pi_{\vec{0},\vec{x}}[i]}$ be the $\vec{u}$-semi-periodic path in $G_\alpha$ originating at $\pi_{\vec{0},\vec{x}}[i]$, and $\pi^{\rightarrow} = \pi_{\vec{0},\vec{x}}[0 \dots i] \cup \pi^{\rightarrow}_{\pi_{\vec{0},\vec{x}}[i]}$.

If $\vec{x} \in \operatorname{dom} \mathcal{F}(\alpha \upharpoonright \pi^{\rightarrow})$ then we are done because $\mathcal{F}(\alpha \upharpoonright \pi^{\rightarrow})$ contains a sub-assembly whose domain is the semi-periodic path $\pi^{\rightarrow}_{\pi_{\vec{0},\vec{x}}[i]}$, which starts in $D_1 \subset D_2$, so assume that $\vec{x} \notin \operatorname{dom} \mathcal{F}(\alpha \upharpoonright \pi^{\rightarrow})$. The path $\pi^{\rightarrow}$ is the base of the comb we will now construct.

Since $\vec{x} \notin \operatorname{dom} \mathcal{F}(\alpha \upharpoonright \pi^{\rightarrow})$ and because there is a simple finite path in $G_\alpha$ from some point $\vec{s}$ on the tail of $\pi^{\rightarrow}$ to $\vec{x}$, Lemma 7.8 tells us that there is a $\vec{v}$-semi-periodic path $\pi$ in $G_\alpha$ originating at $\vec{s}$ with $\pi \cap \pi^{\rightarrow} = \varnothing$. Assume that $\vec{u} \neq z \cdot \vec{v}$ for any $z \in \mathbb{R}$. In this case, let $\pi^{\uparrow} = \pi$. The assembly $\alpha \upharpoonright \pi^{\uparrow}$ is a tooth of some comb with base $\alpha \upharpoonright \pi^{\rightarrow}$. Now define the following assembly.

$$\alpha^* = \alpha \upharpoonright \left( \pi^{\rightarrow} \cup \bigcup_{n \in \mathbb{N}} \left( n \cdot \vec{u} + \pi^{\uparrow} \right) \right).$$

It is clear from the definition that $\alpha^*$ is a comb in $\alpha$ (starting at some point in $D_2$) with respect to $\pi^{\rightarrow}$ and $\pi^{\uparrow}$. By Lemma 7.9, it follows that $\vec{x} \in \mathcal{F}(\alpha^*)$ or $\alpha$ is doubly periodic.

We have shown that, assuming $\vec{u} \neq z \cdot \vec{v}$ for any $z \in \mathbb{R}$, every point $\vec{x} \in X - D_1$ is contained in the finite closure of some comb or periodic path originating at a point in $D_2$ (unless $\alpha$ is doubly periodic). Furthermore, since $D_2$ is finite, there are at most a finite number of combs. The theorem follows by Observation 7.10.

Recall that $\pi$ is an eventually $\vec{v}$-semi-periodic path in $G_\alpha$, and we earlier assumed that $\vec{u} \neq z \cdot \vec{v}$ for any $z \in \mathbb{R}$. Now assume that $\vec{u} = z \cdot \vec{v}$ for some $z \in \mathbb{R}$. Suppose $z < 0$. In this case, let $\pi^{\leftarrow} = \pi$.

If $\vec{x} \in \mathcal{F}(\alpha \restriction (\pi^{\rightarrow} \cup \pi^{\leftarrow}))$, then we are done since $\pi^{\leftarrow}$ originates within $D_2$. If not, then it must be the case that there is a point $\vec{s'}$ on the tail of $\pi^{\leftarrow}$ such that there is a simple finite path from $\vec{s'}$ to $\vec{x}$ in $G_\alpha$. Then Lemma 7.8 tells us that there is an eventually $\vec{w}$-semi-periodic path $\pi'$ in $G_\alpha$ originating at $\vec{s'}$ with $\pi' \cap (\pi^{\leftarrow} \cup \pi^{\rightarrow}) = \varnothing$. Let $\pi^{\nearrow} = \pi'$.

We now have three eventually periodic paths in $G_\alpha$, $\pi^{\leftarrow}$, $\pi^{\rightarrow}$, and $\pi^{\nearrow}$, whose respective tails are all disjoint. Moreover, at least two of these paths must either be "parallel," or two of them are neither parallel nor anti-parallel, and the argument assuming $\vec{u} \neq z \cdot \vec{v}$ for any $z \in \mathbb{R}$ applies. If they are parallel, then by Lemma 7.5, the base $\pi^{\rightarrow}$ forms a periodic path that cuts the plane into two half-planes. Using reasoning that is similar to the above arguments, it is easy to verify that this implies that $\alpha$ is either

1. the finite closure of a (two-way infinite) periodic path bisecting the plane (hence a finite union of semi-periodic sets),
2. (1) unioned with the finite closure of a comb (with a two-way infinite base) covering one of the half planes formed by the periodic bisection, or
3. (1) unioned with two-way infinite combs on both sides of the bisection, each covering one of the half planes.

In each case, it is clear that the set formed is a finite union of semi-doubly periodic sets. □

*Proof.* (*Sketch of Theorem 6.2*) We modify the temperature 2 ("wedge") construction of [3] as follows. In that construction, a row connects to the row to the north, representing the next configuration of $M$, via a double bond at the tile representing the tape head, with a tape head "propagation" tile placed to the north (propagating information about the *next* tape head position). Once the tape head propagation tile is in place, the remainder of the row can fill in, copying the bit from the tile to the south, and with the tile immediately to the left or right of the first tile representing the new tape position. In the revised construction, the bottommost row representing the initial configuration contains strength-1 bonds on the north side, allowing the data on the tape to be copied to the next row, even if the tape head propagation tile has not been placed yet. The tiles copied to the north (and subsequent tiles to the north of them) have bonds to the west and east that do not interact (have strength 0 with each other). The tape head propagation tile is designed to have a strength -1 interaction with the west (resp. east) side if the tape head moves left (resp. right). This "unlocks" the vertical column that may have grown over the new tape head position, removing it so that the tape head tile may now be placed there. The glue strengths on the west (resp. east) side have strength -1 with all glues except for the correct tape head tile, and strength 0 with the tape head tile, which, together with the strength 1 bond from the tile to the south, is enough to allow the tape head tile to bind. Other than the old and new tape head positions, all other tiles in a given row are simply copied from the south with no interaction with neighbors to the west or east. To ensure that the tape head propagation tile is not itself disconnected due to its strength -1 bond with its west or east neighbor, it has a strength 2 bond with the tape head tile to its south. Otherwise, instead of unlocking the vertical column where the tape head is supposed to go, the tape head propagation tile might be removed instead. □